\documentclass[prd,showpacs,amsmath,showkeys,twocolumn,floatfix]{revtex4-1} 
\usepackage{epsfig,dcolumn}
\usepackage{graphicx}
\usepackage[usenames]{color}
\usepackage{graphicx}
\usepackage{bm}

\def\lsim{\mathrel{\rlap{\lower4pt\hbox{\hskip1pt$\sim$}}
    \raise1pt\hbox{$<$}}}
\def\gsim{\mathrel{\rlap{\lower4pt\hbox{\hskip1pt$\sim$}}
    \raise1pt\hbox{$>$}}}
\begin{document}

\title{Schwinger-Dyson equations and disorder}

\author{Adam P.~Szczepaniak$^1$ and Hugo Reinhardt$^2$} 
\affiliation{ $1$ Physics Department and  Center for Exploration of Energy and Matter, Indiana University, Bloomington, IN 47403 USA \\
$2$ Institut f\"ur Theoretische Physik, Auf der Morgenstelle 14, D-72076 T\"ubingen, Germany }

\begin{abstract} 

Using simple models in $D=0+0$ and $D=0+1$ dimensions we construct partition functions and compute two-point correlations. The exact result is compared with saddle-point approximation and solutions of 
Schwinger-Dyson equations. When integrals are dominated by more than one saddle-point we find Schwinger-Dyson equations do not reproduce the correct results unless the action is first transformed into dual variables. 

  \end{abstract} 

\maketitle

\section{Introduction}

We examine applicability of Schwinger-Dyson equations in simple models 
characterized by a nontrivial vacuum.   An infinite set of Schwinger-Dyson equations  (SDE's) represents integral relations between Green's functions that in principle  describe the complete dynamics of the underlying field theory. 
In terms of loop expansion even a single SDE  contains  an infinite series of interaction terms.  For this reason, 
in QCD, where strong interactions between quarks and gluons dominate long range dynamics Schwinger-Dyson
 equations have been extensively used to describe various non-perturbative phenomena;  ranging from 
  confinement and chiral symmetry breaking. to applications in hadron 
phenomenology~\cite{Roberts:1994dr,Alkofer:2000wg,Maris:2003vk}.
 Even when the underlying theory has only a limited  number of elementary interactions the full set of SDE's
 generates a complicated  effective potential.  
 In practical applications any approximation to SDE's 
 eliminates an infinite set of such effective interactions and therefore it is important to access 
  the applicability of any such truncation in QCD phenomenology. 
  A number of investigations in the ultraviolet, and in the more relevant for strong QCD, the infrared region,  have been performed~\cite{Alkofer:2004it,Alkofer:2008tt}.


 There is ample evidence from lattice simulations  that confinement in the QCD vacuum has 
  origin in topology~\cite{Del Debbio:1996mh,Langfeld:1997jx,Engelhardt:1999fd,Greensite:2003bk}. It has been postulated long ago that both 
  confinement and chiral symmetry breaking originate  from instantons. In the case of confinement 
   topological objects like center vortices or magnetic monopole loops 
 percolate through Wilson loops   and lead  to its area law dependence~\cite{Nambu:1974zg,Mandelstam:1974vf,Polyakov:1976fu,'tHooft:1981ht}.  
    As such   a  condensate of magnetic monopoles ought to screen electric flux lines  and produce a finite gluon-gluon correlation length, {\it i.e.} magnetic mass.  The gluon propagator  has been extensively studied using  SDE techniques~\cite{von Smekal:1997vx,Fischer:2008uz,Aguilar:2008xm}   and   it  is therefore worth examining to what extent topological features are manifested in the Green's functions. Examples of such studies in the context of the gluon propagator,  can be found in~\cite{Cornwall:1997ds,Gubarev:1998ss,Chernodub:1999tv,Szczepaniak:2010fe}. 
   
           In this work perform such a study in simple models where SDE solutions can be compared to 
  exact results. In particular with  these models we will be able to address the adequacy 
  of truncated SDE in capturing the underlying, nontrivial properties of the vacuum.  
       In more realistic models with nontrivial topology {\it i.e.}   the Schwinger model~\cite{Adam:1994by,Radozycki:1998xs} or the abelian Higgs model~\cite{Orland:1994qt,Sato:1994vz,Akhmedov:1995mw,Polikarpov:1993cc} Green's functions have been studied using semiclassical approximations  by  introducing dual variables that account for the topological  defects.  Here instead  we will use models in which we can compare SD, semiclassical  and exact results.  In particular we  consider  the following three models that we design to capture, in a much simplified way, 
     some characteristic properties of a topological vacuum of the more sophisticated modes, like the ones mentioned above. 
      In  Sec.~\ref{sec:model-1} we compare truncated SDE results with the exact solution of $D=0+0$ dimensions models   that have either a unique vacuum or degenerate vacua. In 
      Sec.~\ref{sec:model-2} we consider a model with a quasi-periodic vacuum and finally in 
      Sec.~\ref{sec:model-3} we discuss the role of boundary conditions following the example of 
       a particle  on a circle {\it i.e.} a $D=0+1$  field theory. A $D=0+1$ dimensional theory describes  
       a  quantum particle at finite temperature or equivalently classical statistical mechanics of a string. A $D=0+0$ "theory"  may be considered as dimensionally reduced, heavy mass limit of 
       a  $D=0+1$ model.  In the following, however,  we will focus on comparing results of various approximation schemes, including SDE's rather then on their physical interpretation.   Conclusions and outlook are summarized in Sec.~\ref{sec:sum}.



\section{ Unique vs Degenerate Vacuum } 

\label{sec:model-1} 
\begin{figure}
\begin{center}
\includegraphics[width=4 in,,trim=40 0 -60 0,angle=0]{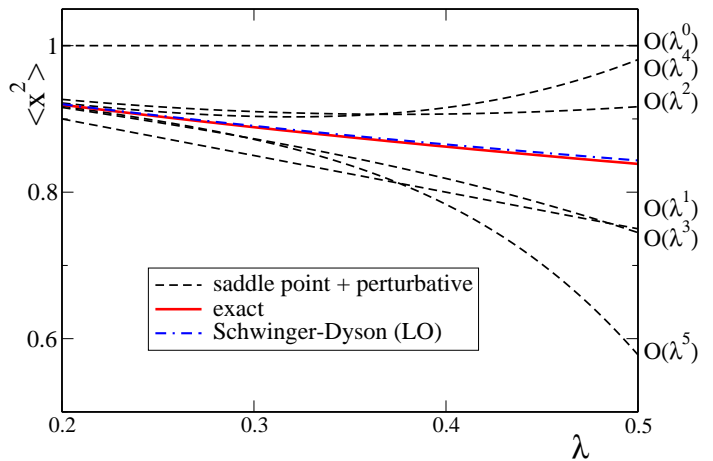}  
\caption{Comparison of  perturbative (dashed lines), Schwinger-Dyson (dashed-dotted line) and exact, numerical evaluation of the two-point correlation (solid line)
 for the  action with a unique classical vacuum ($\epsilon=1$). On this scale
 the NLO SDE solution is indistinguishable  from the exact result. 
\label{fig:mod-1} }
\end{center}
\end{figure}

In $D=0+0$ dimensions the generating functional  becomes a function of a single source variable, $j$ and given by a 
   one-dimensional integral 
\begin{equation} 
Z(j)  = e^{W(j)} = \int dx e^{- S(x) + jx}.  \label{1}
\end{equation} 
For the action  $S(x)$  we take 
\begin{equation} 
S(x)= \epsilon \frac{x^2}{2!} + \lambda \frac{x^4 }{4!} \label{ac}
\end{equation} 
and depending on $\epsilon$ consider both unique and degenerate vacua: 
 if $\epsilon=+1$ the action has a single minimum  while 
 for $\epsilon=-1$ there are two degenerate minima at $x = \pm \sqrt{6/\lambda}$ with the $x=0$ point corresponding to a local maximum.  In $D=0+1$ euclidean dimensions, $x \to x(\tau)$ and  
 the action  $\int_{0}^{\beta} d\tau [\dot x^2(\tau)/2m +S(x(\tau))]$ (with $S (x)$ defined by
(\ref{ac})) describes 
  thermal fluctuations of  a quantum particle, which in the $m \to \infty$ limit reduces to 
    considerations  of integrals as the one given in Eq.~(\ref{1}). 
  We have assumed  that the  integral over $x$ is not restricted 
    {\it i.e.} $x$ runs over the interval $(-\infty,+\infty)$. The semiclassical approximation will be valid in the limit of small coupling $\lambda  \to  0$, as can be easily seen 
      once $x$ is rescaled via, $x \to \bar x = \lambda^{1/2} x$, 
 \begin{equation} 
 S(x) = \frac{1}{\lambda} \left( \epsilon \frac{\bar x^2}{2} + \frac{\bar x^4}{4!} \right). 
 \end{equation} 
In the small-$\lambda$  limit the integral should be well approximated by the contributions from the saddle points. 
 For $\epsilon=+1$ there is one saddle point at $x=0$, which in higher dimensions corresponds 
   to a unique vacuum. For this reason, in the following we refer to saddle points  
  as vacuum contributions, which can be unique, for  $\epsilon=+1$ or multiple as in the case of 
    $\epsilon=-1$ (and in the more general case considered in Sec.~\ref{sec:model-2}). 
 For $\epsilon=+1$ the saddle point approximation is equivalent to the leading order 
standard,   perturbative expansion in powers of $\lambda$.  In particular, the two point correlation 
 \begin{equation} 
 \langle x^2 \rangle  = \frac{d^2 \log Z(j)_{j=0} }{ d j^2}  
 \end{equation} 
  can be easily computed by expanding Eq.~(\ref{1}) in powers of $\lambda$ with the result 
\begin{equation} 
\langle x^2 \rangle  
=   1 
  - \frac{1}{2} \lambda  +  
   \frac{2}{3}  \lambda^2  - \frac{11}{8}  \lambda^3 + \frac{34}{9} \lambda^4 
  -   \frac{619}{48} \lambda^5  +  O( \lambda^6).  \label{sp-1} 
 \end{equation} 
Comparison between the exact, numerical evaluation of  $\langle x^2 \rangle$ and  the  above 
  perturbative series is shown in 
Fig.~\ref{fig:mod-1}. 
As expected, as  $\lambda$ decreases, 
  the accuracy of the saddle point approximation improves, however, even with corrections up 
   to $O(\lambda^5)$   the perturbative expansion is accurate only for 
    very small couplings,  $\lambda \lsim 0.25$. This is an indication of the non-analytical behavior of $Z(j)$ at $\lambda=0$. A similar behavior is also expected in QCD. 
 For larger values of the coupling any reasonable approximation must therefore, 
   at least partially, re-sum the perturbative series to all orders. 
     Since at any order of truncation in the number of effective interactions, Schwinger-Dyson equations do sum up an infinite number of  insertions of $\lambda$ one expects  that a solution of  a truncated set of SDE's  
  will be a better approximation compared to the truncated 
    perturbative expansion of  Eq.~(\ref{sp-1}).  The SD equations follow from the identity 
\begin{equation} 
F' \left[ - \frac{1}{\Gamma^{(2)}(y)}  \frac{d}{d y} + y \right] = - 
\Gamma^{(1)}(y)   \label{sd0}
\end{equation} 
where $\Gamma(y)$ is the  effective action defined by 
\begin{equation} 
 \Gamma(y) = W(j) - j y, \mbox{ with } y = \frac{dW(j)}{dj}. \label{efa} 
\end{equation} 
 With $W(j)$ given by Eq.~(\ref{ac}), Eq.~(\ref{sd0}) leads to the master equation, 
\begin{eqnarray}
  \epsilon y - \frac{\lambda}{6} \left( \frac{\Gamma^{(3)}(y)}{   [\Gamma^{(2)}(y)]^3  } 
 +  \frac{3y}{\Gamma^{(2)}(y)} - y^3 \right) = 
  -\Gamma^{(1)}(y) \nonumber \\ \label{master} 
 \end{eqnarray}
 from which expectation values of any function of $x$, can be generated by taking appropriate  number of derivatives.  In particular the SDE for the  two-point correlation, 
\begin{equation} 
\langle x^2 \rangle  = \frac{d^2W(j)_{j=0}  }{dj^2}=   - \frac{1}{\Gamma^{(2)}_0} 
\end{equation} 
is obtained by taking the first derivative of  Eq.~(\ref{master}) and  setting the source term to zero. This   gives 
\begin{equation}
  \epsilon  + \frac{\lambda}{6} \left( \langle x^2\rangle^3 \Gamma^{(4)}_0 + 3 \langle x^2 \rangle \right) = \frac{1}{\langle x^2 \rangle}.  \label{x22} 
 \end{equation}
   Since $j=0$ implies $y=0$ all terms odd in $y$, in the effective action, Eq.~(\ref{efa}) vanish 
 and the  SD equation for $\Gamma^{(4)}$   is obtained from Eq.~(\ref{master}) by taking two more derivatives of the master equation,  
 \begin{equation}
  -\frac{\Gamma^{(4)}_0}{\lambda}  - 1 = 
+ \frac{3}{2} \langle x^2 \rangle ^4 ( \Gamma^{(4)}_0)^2+ \frac{1}{6} \langle  x^2 \rangle^3 
 \Gamma^{(6)}_0
+ \frac{3}{2} \langle  x^2 \rangle^2  \Gamma^{(4)}_0.
   \label{g4} 
\end{equation} 
Here $ \Gamma^{(n)}_0 = d^n \Gamma(y)_{y=0}/dy^n$ is the dimensionless 
 coupling in the $n$-point vertex of the effective action. 
Similarly, one can derive equations for all higher order vertices, {\it e.g}  $ \Gamma^{(n)}_0$, $n \ge 6$ by taking more derivatives of the master equation. 
Most truncation schemes in applications of SDE's are based on neglecting all but a lowest few vertices.  The lowest order (LO) approximation is obtained by setting 
  $\Gamma^{(4)}_0 = -  \lambda$  {\it i.e.} neglecting dressing of the bare vertex implied by the {\it rhs} of 
  Eq.~(\ref{g4}) as well all higher order vertices  since they 
are generated by higher order  loops.  In the next  to leading order (NLO) one would keep 
 loop 
dressing of the bare vertex, appearing on the {\it rhs} of  Eq.~(\ref{g4}) as terms containing $\Gamma_0^{(4)}$, while continuing to neglect vertices generated by higher order loops, 
   $\Gamma^{(n)}_0 = 0, n \ge 6$.  In LO  the two-point correlator is therefore given by a solution of, 
  \begin{equation} 
\langle x^2 \rangle =  \frac{1}{  \epsilon + \frac{ \lambda}{2} \langle x^2 \rangle 
  - \frac{  \lambda^2}{6}  \langle x^2 \rangle^3 }, \label{lo} 
\end{equation} 
while in the NLO one needs to solve a set of  the two coupled non-linear, algebraic  equations, 
Eq.~(\ref{x22}) and Eq.~(\ref{g4}) with $ \Gamma^{(6)}_0 = 0$. 
It can be easily verified that  the  LO  solution has the following expansion in powers of the coupling constant $ \lambda$
 \begin{equation} 
\langle x^2 \rangle =  1  - \frac{1}{2} \lambda + \frac{2}{3} \lambda^2 - \frac{9}{8}  \lambda^3  + O( \lambda^4) 
\end{equation} 
{\it i.e.} it agrees with the exact result only  up to the second order, while  for the  NLO solution one finds, 
\begin{equation} 
\langle x^2 \rangle =  1   - \frac{1}{2}  \lambda + \frac{2}{3}  \lambda^2  - \frac{11}{8}  \lambda^3  + \frac{34}{9}  \lambda^4  - \frac{599}{48}  \lambda^5  +  O( \lambda^6) 
\end{equation} 
which agrees  with Eq.~(\ref{sp-1}) up to the fourth order.  In Fig.~\ref{fig:mod-1} we also compare the saddle-point  approximation   of Eq.~(\ref{sp-1}) with the LO and NLO solution of SDE's for $\epsilon=+1$. Clearly the SDE equations 
result in the two-point correlation that is significantly more accurate then the saddle point approximation (aka perturbation theory)   for  large values of $\lambda$. 
It is worth noting that this not because the solution of the SDE's and the perturbative series match to some 
  high order in $\lambda$; clearly the perturbative expansion is quite inaccurate except for very small  $\lambda$. 
The agreement between solutions of SDE's and the exact result 
 originates  from the effective all order re-summation of the perturbative series generated by the nonlinear SDE's.  Indeed comparing the coefficients of the first few terms in perturbative   expansion beyond the order where SED's match the exact expansion,  {\it e.g.} $619/48$ vs $599/48$  
  one finds  the difference to be only  a  few percent. 

\begin{figure}
\begin{center}
\includegraphics[width=4 in,,trim=10 0 -60 0,angle=0]{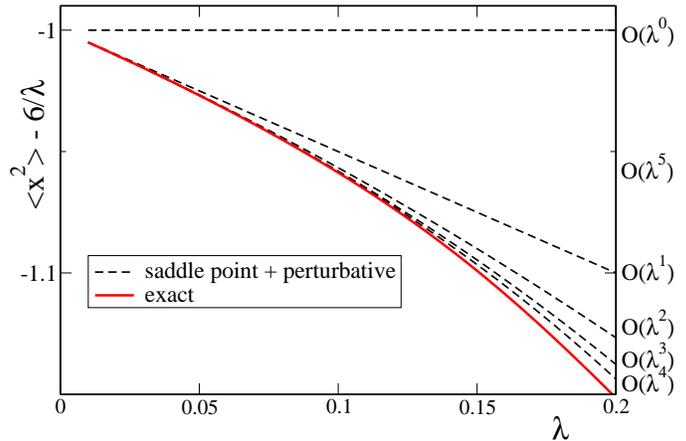}  
\caption{  Saddle-point approximation  computed to a varying order in $\lambda$ (dashed lines) compared 
 with the exact result for $\epsilon=1$ obtained by numerical integration (solid line).   The $O(\lambda^4)$ perturbative results is indistinguishable form the exact result.  
 For easier comparison we subtracted the leading, 
  $6/\lambda$ term from $\langle x^2 \rangle$. On this scale the solution of SD equation tends to  
    $-6/\lambda \to -\infty$  as  $\lambda \to 0$. 
  \label{fig:mod-2} }
\end{center}
\end{figure}

For  strong coupling, $\lambda > 1$, however, the SDE's truncated so that  higher vertices (loops) are removed, fail. This is because higher order vertices in the effective action, $\Gamma^{(n)}_0$ grow with $\lambda$. The perturbative  expansion near the $x=0$ saddle point can nevertheless be set up  by  expanding the integrand in Eq.~(\ref{1}) in terms of the 
 "kinetic term", $\epsilon x^2/2$, instead of the interaction term,  {\it i.e.} by expanding in powers of $1/\sqrt{\lambda}$. This leads to 
 \begin{equation} 
 \langle x^2 \rangle = \frac{2\sqrt{3}\Gamma^2(\frac{3}{4})}{\pi \sqrt{\lambda}}
   \left(1 + 
 \sum_{n=1} \frac{c_n}{(\sqrt{\lambda})^n}  \right) 
 \end{equation} 
where all higher order coefficients, $c_n$  can be computed analytically.  It then follows  from 
Eqs.~(\ref{x22}),~(\ref{g4}) that for large-$\lambda$ 
\begin{equation} 
\frac{\Gamma^{(4)}_0}{\lambda}  \approx - 0.2960   + O(\frac{1}{\sqrt{\lambda}}), \;
\frac{ \Gamma^{(6)}_0}{\lambda^{3/2}}  \approx -0.6276   + O(\frac{1}{\sqrt{\lambda}}).
\end{equation} 
{\it i.e.} there is no suppression of higher order vertices, $\Gamma^{(n)}/\lambda^{n/4} = O(1)$.
For  $\epsilon=-1$, SD equations  truncated at any finite order  cannot reproduce the exact result. This is because vertices in the SDE's generated from the master equation  
  originate from  expansion around $x=0$,  which is a local maximum and a metastable sate in higher dimensions. 
 For $\epsilon =-1$, $\langle x^2 \rangle$ is still positive  and as a function of $\lambda$ it is non-analytical  at $\lambda = 0$, where it has a pole, 
  while the LO solution Eq.~(\ref{lo})  is analytical and for  $\lambda = 0$  gives $\langle x^2\rangle_{LO} = -1$! 
   The action has two minima and in the saddle point approximation the integral in Eq.~(\ref{1}) is approximated  by a sum of gaussian fluctuations around each of them  with the difference between the full action and gaussian approximation treated as perturbation. This leads to 
    \begin{equation} 
 \langle x^2 \rangle = \frac{6}{\lambda} - 1 - \frac{1}{2} \lambda 
  - \frac{2}{3} \lambda^2 - \frac{11}{8}\lambda^3 - \frac{34}{9}\lambda^4 - \frac{619}{48} \lambda^5 + O(\lambda^6) 
  \end{equation} 
and is compared to the exact,  numerical result in Fig.~\ref{fig:mod-2} 

\section{ Multiple quasi-degenerate vacua }
\label{sec:model-2} 

In QCD large gauge transformations are not constrained by the Gauss'  law and result in
 topologically disconnected field configurations~\cite{inst1,inst2,Reinhardt:1997rma,Ford:1998bt,Jahn:1998nw}. In the semiclassical
 approximation  these configurations  correspond to degenerate classical vacua 
  and the true vacuum state is a linear superposition of these vacua. 
To access the applicability of SDE equations in  the case of such multiple saddle points, 
 we again make a simple model for the action in $D=0+0$.    
 Since there is no tunneling without 
 the "time" direction, it is possible that  SDE's equations in the realistic case with  
  instantons are more accurate  then the one ones in the  no-time model discussed here.  
 
\begin{figure}
\begin{center}
\includegraphics[width=3.5 in,angle=0]{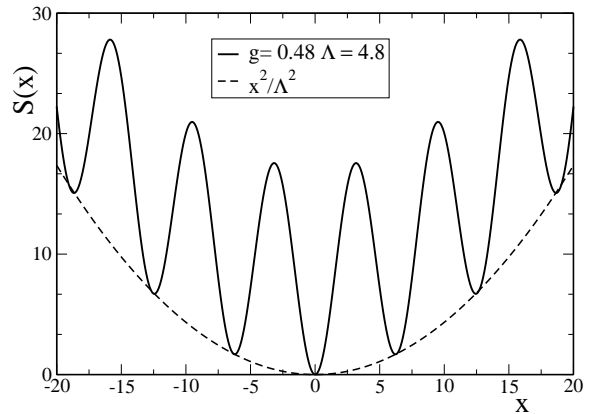}  
\caption{ A typical action used here in a model study of multiple, quasi-degenerate vacua.
  For large $\Lambda$ and/or small $g$ the 
 the number of minima contributing to the partition function  is proportional to $\Lambda^2/g^2$. For the case shown in this figure, besides the central minimum there are additional 
  $N=32$  local minima. 
\label{fig:wf}}
\end{center}
\end{figure}

\begin{figure}
\begin{center}
\includegraphics[width=3.5 in,angle=0]{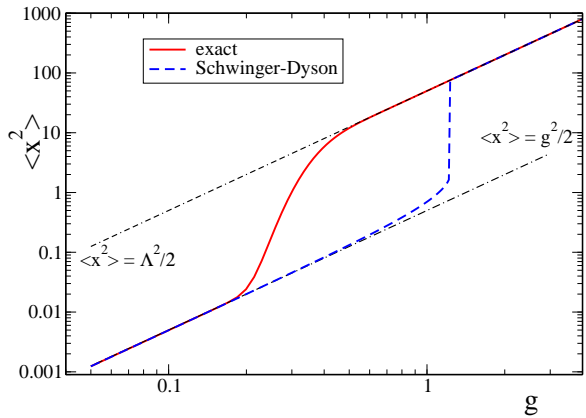}  
\caption{ Comparison between exact  (solid line) and Schwinger-Dyson,  Eq.~(\ref{gap}) approximation (dashed line)  to $\langle x^2 \rangle$ for the action given by Eq.~(\ref{ac2}).
 We used $g^2/\Lambda^2 = 0.01$ ({\it cf.} Fig.\ref{fig:wf}). The dashed-dotted lines correspond to the two limits given by Eqs.~(\ref{low}),~(\ref{dd}), were, as discussed in the text Schwinger-Dyson approximation  becomes exact. For the range of couplings shown the saddle point approximation of Eq.~(\ref{sd-mod2})  is indistinguishable from the exact result and involves summation over  $N=33$ saddle points. 
\label{exact-vs-ds}}
\end{center}
\end{figure}

 To mimic the effect of multiple vacua in the  partition function we consider the following action 
\begin{equation} 
S(x) =  \frac{1}{g^2}  \left[ 2 ( 1 - \cos(x))  + \frac{x^2}{\Lambda^2}  \right]. \label{ac2} 
\end{equation} 
For simplicity, to be able to compare with  SDE's  we define 
  $x$ in the  $(-\infty, +\infty)$ interval. The multiple saddle points originate from the $\cos$ term the role of the second term on {\it rhs} in Eq.~\ref{ac2} is 
 to make the integral in Eq.~(\ref{1}) well defined. Because of this term the minima of  $S(x)$ are not really degenerate, and their relative contribution  depends on $\Lambda$. 
 For weak coupling, $ g < 1$   the expectation value of $\langle x^2 \rangle$ is determined in the saddle point approximation  by  the number of minima of the action. For fixed, finite  $\Lambda$,  $\langle x^2 \rangle \to 0$ when 
 $g \to 0 $ since in this case the term $x^2/(g^2\Lambda^2)$ is large for all minima of $\cos x$ except the one at  $x=0$.   In this limit, therefore,  the integral  becomes dominated by the single minimum at $x=0$  and one finds. 
 \begin{equation} 
 \langle x^2 \rangle \to \frac{\Lambda^2 g^2}{2 (\Lambda^2 + g^2)} \to \frac{g^2}{2}.  \label{low} 
 \end{equation} 
Since the problem becomes effectively that of a single vacuum one expects  
  the SDE  to yield a similar result. This indeed is the case as shown  in Fig.~\ref{exact-vs-ds}.  As $\Lambda$ increases for fixed $g<1$ the contribution from the  minima of  $\cos x$ at $x = 2\pi n$, $|n| > 0$   are  no longer   suppressed and in the limit $\Lambda \to \infty$ one obtains 
 \begin{equation} 
 \langle x^2 \rangle \to \frac{\Lambda^2}{2}.  \label{dd} 
 \end{equation} 
This is an interesting limit, since even though there are multiple vacua contributing to the partition function integral their contribution is approximately equal to that of a broad  minimum given by 
 $x^2/\Lambda^2$ ({\it c.f.} 
Fig.~\ref{fig:wf}),  and it is this gaussian distribution that results in  Eq.~(\ref{dd}).

The SD equations for the action of Eq.~(\ref{ac2}) are derived from the 
operator identity 
\begin{equation} 
\frac{2}{g^2} \sin\left[ -\frac{1}{\Gamma^{(2)}(y)} \frac{d}{dy} + y \right] 
  + \frac{2y}{\Lambda^2} 
    = -\Gamma^{(1)}(y). \label{sd2} 
    \end{equation} 
In particular for the two-point correlation in the LO approximation, $\Gamma^{(2)}(y)  = \Gamma^{(2)}_0$, it yields, 
 \begin{equation} 
\frac{2}{g^2}\frac{d}{dy}  \sin \left[ -\frac{1}{\Gamma^{(2)}_0} \frac{d}{dy} + y \right]  + \frac{2}
{\Lambda^2} = - \Gamma^{(2)}_0. \label{m2sde} 
\end{equation} 
The non-linear, algebraic equation for $\langle x^2 \rangle = - 1/\Gamma^{(2)}_0$ can be derived from the above by noticing that  the argument of 
the sine can be written as an expectation value of  ladder  operators 
\begin{equation} 
\frac{d}{dy}  \sin \left[ -\frac{1}{\Gamma^{(2)}_0} \frac{d}{dy} + y \right]  = 
\sqrt{\langle x^2 \rangle} \langle 0|  a \sin \left[ \frac{a+a^{\dag}}{\sqrt{\langle x^2\rangle}} \right]| 0 \rangle  
\end{equation} 
with $a \equiv  1/\sqrt{\langle x^2 \rangle} d/dy$ and $a^{\dag} \equiv \sqrt{\langle x^2 \rangle} y $.
 It  finally leads to the following gap equation 
\begin{equation} 
\frac{2}{g^2} e^{-\frac{\langle x^2 \rangle}{2} } + \frac{2}{\Lambda^2} = \frac{1}{\langle x^2 \rangle}.  
\label{gap}  
\end{equation} 
It can be verified that for finite $\Lambda$ in the $g \to 0$ limit the solution of Eq.~(\ref{gap}) 
agrees with Eq.~(\ref{low}) while in the limit $\Lambda \to \infty$ it agrees with Eq.~(\ref{dd}) 
In the former case,  as discussed above, the SD equation reproduces 
 the result of the problem with a unique vacuum. For large-$\Lambda$ on the other hand,  it indeed follows from Eq.~(\ref{gap}) that 
  even though multiple saddle points contribute  the net effect  is  equivalent to that of a  single minimum with a large width $\sim \Lambda$ 
  leading to solution for  $\langle x^2\rangle$ that  agrees with Eq.~(\ref{dd}), 
  as shown in Fig.~\ref{exact-vs-ds}. 
  
\begin{figure}
\begin{center}
\hspace{5cm}\includegraphics[width=3.5 in,angle=0]{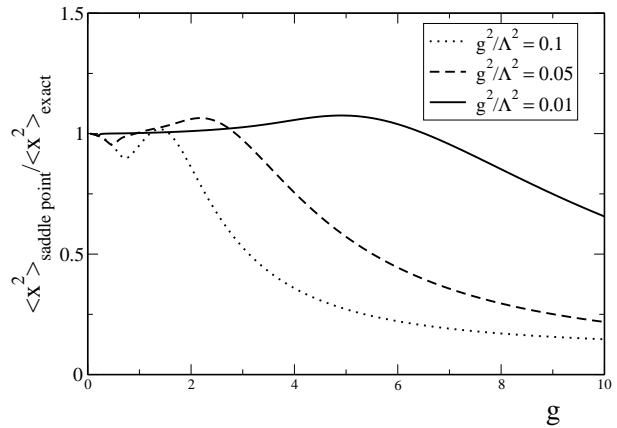}
\caption{ Comparison between exact, numerical computation  and the saddle point approximation 
 to  $\langle x^2 \rangle$ for different values of $g^2/\Lambda^2$.  As $\Lambda$ increases for fixed $g$ so does the number of saddle points. The three cases shown correspond to $N=3,7,33$ points respectively ({\it cf}. Eq.~(\ref{sd-mod2})). As  $\Lambda$ increases, saddle points become degenerate and Eq.~\ref{sd-mod2} becomes 
  an increasingly better approximation. 
  \label{saddle-vs-exact}}
\end{center}
\end{figure}

 Outside of the two limits, several local minima of he action contribute to the partition function and the LO SD equation fails. This is seen in Fig.~\ref{exact-vs-ds} for $g$ in the mid-range, $0.2 \lsim g \lsim 1$. In this range, where multiple minima contribute  
  one expects summing over integrals around all  saddle points is a better approximation to the generating functional compared to the SDE. 
   In this,  saddle point approximation 
 \begin{equation} 
 \langle x^2 \rangle = \frac{\sum_{i=1}^N 
  ( x_i^2 + \frac{1}{2\omega_i})
 \frac{e^{-S(x_i)}}{\sqrt{\omega_i}}  
 }{ 
 \sum_{i=1}^N \frac{e^{-S(x_i)}}{\sqrt{\omega_i}} }  \label{sd-mod2} 
 \end{equation}   
 and comparison with the exact, numerical result is shown in Fig.~\ref{saddle-vs-exact}. 
  Indeed the saddle point approximation works while  SDE  fails in this range of couplings.

 \section{ Particle on a circle } 
 \label{sec:model-3} 
\begin{figure}[h]
\begin{center}
\hspace{3.5cm}\includegraphics[width=3.5 in,angle=0]{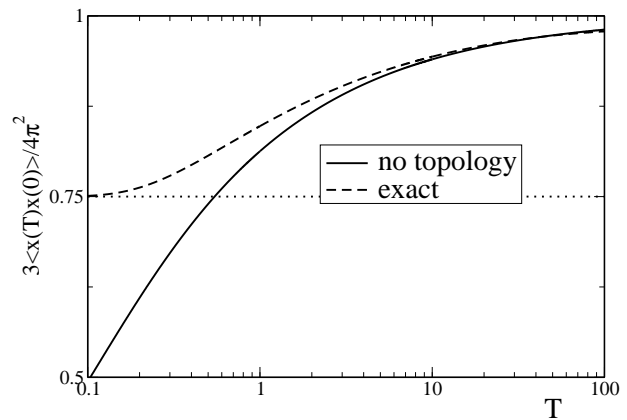}
\caption{ Comparison between exact (dashed  line) and  approximate evaluation of the correlation   function as a function of temperature. In the approximation 
  Eq.~(\ref{clas}) only the $q=0$ term is retained. 
\label{fig:poc}}
\end{center}
\end{figure}

In QCD the domain of gauge fixed gluon field, {\it  i.e.} the fundamental modular region 
 may be non-flat  with a non-trivial measure specified by the Fadeev-Popov determinant~\cite{Gribov:1977wm}. 
  This highly complicates the Schwinger-Dyson formalism, which needs to take into account 
   the boundary of fundamental modular region or the Gribov horizon~\cite{Zwanziger:1989mf}. 
  In the following example we  investigate what  happens  in a model in which  the dynamical variable has a non-trivial boundary under an approximation when this boundary is ignored. 
  For this purpose we consider quantum mechanics of a particle on a circle~\cite{poc} {\it i.e.} $D=0+1$ dimensional field theory.  The variable $x(t)$ now describes location of the particle on a unit circle   $0\le x(t)  <  2\pi$ as a function of time and for the Hamiltonian we choose the free kinetic energy for a particle of unit mass  $2H = p^2 = - \partial^2_x $. Since the manifold is compact, the wave function must satisfy the boundary condition $\psi(x) = \exp(i\theta) \psi(x + 2\pi)$ and in the following we take $\theta=0$.  The normalized  eigenvectors of the Hamiltonian are spanned by $\psi_m(x) = \exp(imx)/\sqrt{2\pi}$ with integer $m$ and the corresponding energies are 
   $E_m = m^2/2$.
    The vacuum expectation value at 
 the  euclidean time ($t = -i \beta$), {\it i.e.} temperature-dependent  correlation function is then given by
     \begin{equation} 
\langle x(\beta) x(0) \rangle = \langle 0| x e^{-\beta H} x |0 \rangle = \pi^2 + \sum_{m\ne 0} \frac{e^{-\beta\frac{m^2}{2}}}{m^2}.  \label{quantum} 
\end{equation} 
At low temperatures, $T= 1/\beta \to 0$ the correlation function is dominated by the lowest  energy quantum sate  and   $\langle x(\infty) x(0) \rangle \to \pi^2$. In this case the restriction that  $x$ 
 be on a circle  is  important.   At high temperatures, however,  the system becomes  semiclassical,  and the particulars of the topology of the quantum system should become irrelevant. Also 
  in this  limit expectation values should be well approximated by contributions from  
  small amplitude fluctuations around  solutions of  the  
classical  equation of motion.  In this case 
 truncated    SD equations should also be a good approximation. 
 In our simple example we assume no interaction thus the SD equation and the semiclassical,  
  saddle point approximation give the same results and 
 both of  them pertain to a formulation of 
 the problem in terms of the variable dual to quantum number  $m$
 ~\cite{Banks:1977cc}.  This variable is just is the  classical coordinate
   $x$ and the duality transformation $m \leftrightarrow x$   is given by 
   \begin{equation} 
   e^{-\beta  \frac{m^2}{2}} =    \int_{-\infty}^{\infty} \frac{dx}{\sqrt{2\pi\beta}}  e^{-\frac{x^2}{2\beta} -   i m x}  
   \end{equation} 
 leading to 
   \begin{equation} 
   \langle x(\beta) x(0)\rangle = \int_0^{2\pi} \frac{dx'}{2\pi} \frac{dx}{2\pi} 
   x' \langle x',\beta|x,0\rangle x
   \end{equation} 
   where 
   \begin{equation} 
   \langle x',\beta|x,0\rangle =  \sqrt{\frac{2\pi}{\beta}}  \sum_{q\in N}  \exp\left( - \frac{(x' -x + 2\pi q)^2}{2\beta}\right).  \label{clas} 
   \end{equation}  
 At finite temperature, $q$ counts the number of times the particle wraps around the circle. 
   The duality  between $m$ and $x$ is clear;  at high temperature, $\beta \to 0 $ and 
   $x$ is well defined, (while $m$ is not) since 
   \begin{equation} 
    \langle x',\beta|x,0\rangle \to 2\pi \delta(x' - x)
    \end{equation}
 with only the term with $q=0$ contributing. It then immediately follows that, 
    $\langle x(0) x(0)\rangle = 4\pi^2/3$. In the high temperature limit,  
   the quantum variable $m$  is not well defined, {\it i.e.} in the sum in Eq.~(\ref{quantum}) an infinite number of terms   contribute to give $4\pi^2/3$. At low temperatures, on the other hand, the system is quantum. As $\beta \to \infty$,  $m$  becomes well defined, $m \to 0$ (while $x$ is not)   and 
    $\langle x(\infty)x(0) \rangle \to \pi^2$. 
To obtain the  low temperature limit of the correlation function using the classical representation of  Eq.~(\ref{clas})  it is necessary to integrate over the entire range of  $x$ and $x'$. 
    Furthermore  it is necessary to allow $x$ (or $x'$) to wrap around the circle an arbitrary number of times {\it i.e.} sum over all $q\in N$. 
        The approximation in which the topology of the boundary is ignored corresponds to retaining only the  $q=0$ term in the sum 
      in Eq.~(\ref{clas}),  which, as follows from the discussion above, should work fine in the high temperature limit  but fail at low temperatures. This is  shown in Fig.~\ref{fig:poc}.

   


\section{ Summary} 
\label{sec:sum} 
Phenomenological applications of Schwinger-Dyson equations require truncations in the number of retained effective interactions. It is important to access the reliability of such truncation in QCD where Green's functions are not necessarily dominated by small 
fluctuations around the perturbative vacuum. In particular,  confinement is expected to be related to 
  topologically non-trivial field configuraiton like magnetic monoples and vortices that lie on 
the Gribov horizon \cite{Greensite:2004ke} and produce tunneling between the degenerate,  local minima of the action.   If the effective potential for gauge degrees of freedom were known 
 in principle one could compare the SDE approach with other, {\it i.e.} semiclassical approximations, 
 (as can be done for example in the case of compact QED~\cite{Drell:1978hr}). 
 Here we have done such comparison in simple models to illustrate strengths and  potential limitations of the SDE approach. In particular we have shown that if  the partition function integral is 
  dominated by a single  semiclassical  configuration, (one saddle point of the action), even  the one-loop  SD equation gives a much better approximation   
    for a two-point correlation than perturbative  
   series truncated at some high order.  In the case of multiple-saddle points, however, we have shown that
 the semiclassical approximation, when applicable is more reliable while the SDE approach may not necessarily be 
capturing the correct physics. In such cases SDE's can, however be successful when formulated in dual variables.  
  Clearly the models studied  here are very naive and more realistic problems need to be considered before definite
 conclusions about applicability of truncated SD equations to describe physics of confinement are made. 
   
\vspace{5cm}   
\acknowledgments{This work was  supported in part by the  US Department of Energy grant under
contract DE-FG0287ER40365  and by DFG under contract DFG-Re856/6-3. }

\end{document}